\documentclass[aps,prb,superscriptaddress,twocolumn]{revtex4-2}

\usepackage{graphicx}
\usepackage{dcolumn}
\usepackage{bm}
\usepackage{amsmath}
\usepackage[dvipsnames]{xcolor}
\usepackage[margin=0.75in]{geometry}
\usepackage[normalem]{ulem}
\usepackage[colorlinks=true, linkcolor=blue, citecolor=blue, urlcolor=blue]{hyperref}
\usepackage{xcolor}

\usepackage{multirow}
\usepackage{booktabs}

\begin{document}

\title{Local magnetic order in vacancy-disrupted spin ice Ho$_2$TiO$_5$}

\author{Raju Baral}
\email{baralr@ornl.gov}
\affiliation{Neutron Scattering Division, Oak Ridge National Laboratory, Oak Ridge, Tennessee, 37831, USA}

\author{Haidong Zhou}
\affiliation{Department of Physics and Astronomy, University of Tennessee, Knoxville, Tennessee, 37996, USA}

\author{Qiang Zhang}
\affiliation{Neutron Scattering Division, Oak Ridge National Laboratory, Oak Ridge, Tennessee, 37831, USA}

\author{Benjamin A. Frandsen}
\affiliation{Department of Physics and Astronomy, Brigham Young University, Provo, Utah, 84602, USA}

\author{Stuart Calder}
\email{caldersa@ornl.gov}
\affiliation{Neutron Scattering Division, Oak Ridge National Laboratory, Oak Ridge, Tennessee, 37831, USA}

\begin{abstract}

We investigate how local magnetic correlations evolve when the classical pyrochlore spin-ice Ho$_2$Ti$_2$O$_7$ is transformed into the partially disordered stuffed compound Ho$_2$TiO$_5$. Neutron scattering measurements were analyzed using real-space magnetic pair distribution function, reciprocal-space reverse Monte Carlo, and half-polarized neutron powder diffraction methods to connect the average crystal structure with local magnetic correlations. Both compounds retain long-range $Fd\bar{3}m$ symmetry, but in Ho$_2$TiO$_5$ this average structure distorts locally through Ho--O, Ti--O and O--O bond-length disorder associated with partial Ho/Ti occupancy. Despite this disorder, half-polarized neutron powder diffraction shows that the Ho moments retain local $\langle111\rangle$ Ising anisotropy and form spin-in/spin-out configurations on the tetrahedral network. In Ho$_2$Ti$_2$O$_7$, real- and reciprocal-space analyses reveal a nearly ideal two-in/two-out spin-ice state at 0.3~K, with ${\sim}95\%$ of tetrahedra satisfying the ice rule, followed by progressive thermal disordering on warming. In Ho$_2$TiO$_5$, most tetrahedra are magnetically incomplete because some tetrahedral vertices are occupied by nonmagnetic Ti rather than Ho; nevertheless, the dominant incomplete configurations are 2-in/1-out and 1-in/2-out, which are locally compatible with the ice rule if the missing Ho spin is restored. The single-ion $\langle111\rangle$ Ising anisotropy is therefore retained throughout, while the collective ice ordering is strongly disrupted: Ho$_2$TiO$_5$ is a vacancy-disrupted spin ice in which a local ice-rule tendency persists on a topologically incomplete magnetic network.

\end{abstract}

\maketitle

\section{Introduction}

Geometrically frustrated materials have been a central theme in condensed matter physics for decades, attracting intense interest due to their rich magnetic phenomena, novel ground states, and strong quantum fluctuations \cite{RevModPhys.82.53}. Geometrical magnetic frustration arises when the geometry of the crystal prevents all the competing spin interactions from being simultaneously satisfied, leading to competing ground states and often suppression of long range order while preserving local magnetic correlations. Pyrochlore titanates ($A_2$Ti$_2$O$_7$, $A$ = Rare earth ion) have emerged as one of the most fertile families of three-dimensional frustrated magnets. In this structure, the magnetic A-site cations sit on the vertices of a network of intrinsically frustrated corner-sharing tetrahedra. One prominent member Ho$_2$Ti$_2$O$_7$ is a canonical spin-ice material, where each tetrahedron adopts a “2-in/2-out” spin configuration, that does not undergo long range order down to the lowest measured temperatures~\cite{Harris_PRL_Ho227_1997, Bramwell_1998_Ho227, Bramwell_2001_Ho227}. The magnetic Ho$^{3+}$ ion's strong crystal-electric-field environment at the $A$ site enforces a pronounced Ising anisotropy, locking each moment to its local $\langle 111 \rangle$ direction so that it can point only toward or away from the center of a given tetrahedron~\cite{Harris_PRL_Ho227_1997}. As a result, the lowest-energy states satisfy the spin-ice rule.

\begin{figure}[tbh]
    \centering
    \includegraphics[width=70mm]{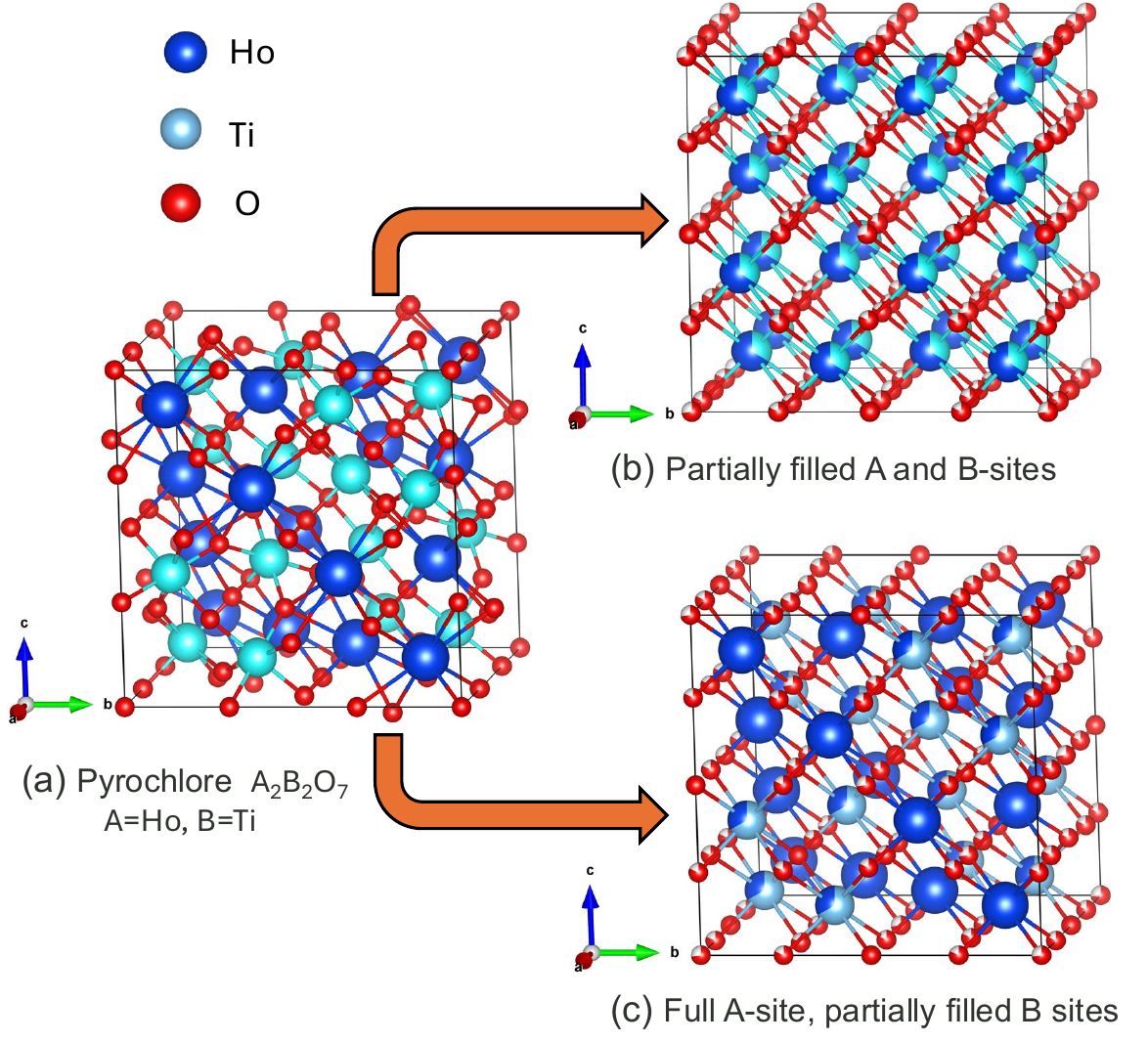}
    \caption{\label{fig:struc} Crystal structure models for pyrochlore spin-ice A$_2$B$_2$O$_7$ and stuffed spin-ice A$_2$BO$_5$, where A = Ho and B = Ti. (a) Pyrochlore Ho$_2$Ti$_2$O$_7$ with fully occupied $A$-site Ho and $B$-site Ti sublattices. (b,c) Idealized average-structure descriptions of Ho$_2$TiO$_5$ showing two possible cation-occupancy models. Panel (b) shows partial Ho occupancy on both the $A$ and $B$ sites, with approximately two-thirds occupancy on each sublattice. Panel (c) shows a fully occupied $A$ site and a partially occupied $B$ site.}
\end{figure}

The spin-ice behavior of Ho$_2$Ti$_2$O$_7$ is well established; however, a fundamentally important and largely unexplored question is what happens to this ground state when structural disorder is introduced into the lattice. Because spin-ice is governed by local constraints on the Ho moments, a natural question is therefore how robust this behavior remains when the tetrahedra are no longer magnetically complete. One route to create this scenario is through cation stuffing, where additional Ho ions occupy nominally nonmagnetic Ti sites and drive Ho/Ti mixing on both the $A$ and $B$ sublattices (see Fig.~\ref{fig:struc}). This stuffing pushes the composition beyond the stoichiometric limit toward Ho$_2$(Ti$_{1.33}$Ho$_{0.67}$)O$_{6.67}$, or simply Ho$_2$TiO$_5$. The substitution changes both the connectivity of the magnetic network and the local crystal-field environment, potentially disrupting the balance of interactions that stabilizes the ice-rule ground state.

Stuffed spin ices therefore provide a route to test the robustness of ice-rule correlations against chemical disorder and magnetic dilution. Previous work on stuffed pyrochlores has shown that the average cubic symmetry can persist over a wide stuffing range, even as the local structure and magnetic response are strongly modified \cite{Lau2006Nature, Aldus_2013, PhysRevB.76.054430, LAU20063126, LAU200845}. Ho$_2$TiO$_5$ represents an extreme limit of this tuning, where the average structure remains pyrochlore-like, but the local Ho/Ti network produces a large population of magnetically incomplete tetrahedra. Here, we use “vacancy” to denote a magnetic vacancy on the Ho spin network, corresponding structurally to a site occupied by nonmagnetic Ti rather than magnetic Ho.

The structural consequences of this substitution add a further layer of complexity. Depending on the synthesis conditions, the excess Ho can be accommodated in two distinct ways. One possibility is a fully occupied $A$ site with partial occupancy of the $B$ site. Another is simultaneous partial occupancy of both the $A$ and $B$ sites, with approximately two-thirds filling on each site (Fig.~\ref{fig:struc}) \cite{PhysRevB.76.054430, LAU20063126, LAU200845}. These two configurations are not equivalent from a magnetic perspective, as they differ in how the Ho moments are distributed across the lattice and how vacancies interrupt the tetrahedral spin network.

Our analysis integrates a suite of complementary approaches: atomic/magnetic pair distribution function (PDF) analysis, magnetic reverse Monte Carlo (RMC) modeling, and half-polarized neutron powder diffraction (pNPD) to interrogate both the average and local atomic structures, as well as the short-range magnetic correlations, in Ho$_2$Ti$_2$O$_7$ and Ho$_2$TiO$_5$. Refinement of the average structure of Ho$_2$TiO$_5$ confirms retention of the $Fd\bar{3}m$ space group symmetry; however, systematic misfits in the low-$r$ region of the PDF highlight the presence of local structural deviations that are not captured by long-range crystallographic descriptions. By analyzing the magnetic pair distribution function (mPDF)~\cite{frandsen2015mPDFMnO,Frandsen:mPDF_paper_2014} and employing a magnetic reverse Monte-carlo (RMC) program \textsc{Spinvert}~\cite{Paddison_2013} in both real and reciprocal space, we directly resolve the short-range magnetic correlations in both compounds. In Ho$_2$Ti$_2$O$_7$, the spin configurations satisfy the ice rule at low temperature, with increasing temperature leading to a progressively larger fraction of tetrahedra violating this constraint. In the stuffed spin ice Ho$_2$TiO$_5$, over 80\% of tetrahedra are topologically magnetically incomplete, a direct structural consequence of the partial, $2/3$-fractional occupancy of Ho$^{3+}$ across both the $A$- and $B$-site sublattices, and among these, the dominant spin configurations are 2-in-1-out and 1-in-2-out arrangements. The remaining $\sim$20\% of complete tetrahedra exhibit either 2-in-2-out or 3-in-1-out/1-in-3-out configurations. This intrinsic site disorder constitutes a defining feature of stuffed spin-ice and underpins the unconventional magnetic behavior observed in this system. Furthermore, our half-polarized neutron powder diffraction data (pNPD) analysis confirms that the spins are constrained to point either towards or away from the center of the tetrahedra. Consequently, the single-ion $\langle111\rangle$ Ising anisotropy that underpins spin-ice physics is retained on both sublattices, even as the topologically incomplete network strongly disrupts the collective two-in/two-out ordering, leaving a local ice-rule tendency rather than the near-ideal order of Ho$_2$Ti$_2$O$_7$.

\section{Experimental section}

Powder samples of Ho$_2$TiO$_5$ and Ho$_2$Ti$_2$O$_7$   were synthesized using stoichiometric amounts of dried Ho$_2$O$_3$ and TiO$_2$. The mixed powders were subsequently annealed in air at 1000$^\circ$C, 1200$^\circ$C and 1450$^\circ$C for 20 hours at each temperature.

Neutron total scattering patterns of Ho$_2$Ti$_2$O$_7$ and Ho$_2$TiO$_5$  were collected on the HB-2A powder diffractometer ~\cite{mPDF_HB2A_Baral} at the High Flux Isotope Reactor (HFIR) at Oak Ridge National Laboratory (ORNL). A vertically focusing germanium monochromator was used to select the wavelengths of 1.12 $\mathrm{\AA}$ from the Ge(117) reflection. The total neutron scattering data for both samples were collected at temperatures 0.3, 1, 2, 4, 10, 20, 40, and 50 K, with a collection time of 6 hours for each temperature point. The collected data were then Fourier transformed with $Q_{\mathrm{max}}= 10~ \mathrm{\AA^{-1}}$ to get the pair distribution function (PDF) data. 

Half-polarized neutron powder diffraction (pNPD) data were also collected on HB-2A using the 2.41 $\mathrm{\AA}$ wavelength from the Ge(113) monochromator reflection. For this data collection a polarizing supermirror V-cavity was placed in the incident beam to define a single neutron polarization state. Data were collected with the neutron spin polarization up or down with the further use of a Mezei guide-field flipper \cite{mPDF_HB2A_Baral}. The sample was made into pellets to avoid grain reorientation and placed in a vertical magnetic field.

Additional neutron total scattering patterns of Ho$_2$TiO$_5$ were collected at 10 and 300 K on the POWGEN powder diffractometer at the Spallation Neutron Source (SNS) at ORNL. An automatic sample changer was used as the sample environment, covering the temperature range from 10 to 300 K. Approximately 5 g of powder was loaded into an 8 mm diameter vanadium PAC can and sealed with helium exchange gas to ensure good thermal conductivity. Data were collected using a neutron frame with center wavelengths of 0.8 \AA, providing d-spacing coverage of 0.1–8 \AA. All the data were reduced by subtracting the signal from the empty PAC container, with normalization performed using the difference between the vanadium standard and the empty instrument background. The collected data were Fourier transformed with $Q_{\mathrm{max}}= 30~ \mathrm{\AA^{-1}}$ to obtain the pair distribution function (PDF). 

Atomic PDF analysis of the data was performed using the open source Python package \textit{ DiffPy-CMI}~\cite{Juhas_diffpy_cmi}, whereas mPDF analysis was performed using the open source Python package \textit{diffpy.mpdf}~\cite{frandsen2022diffpympdf,frandsen2015mPDFMnO,Frandsen:mPDF_paper_2014}. Magnetic RMC modeling was performed using the \textsc{Spinvert} software \cite{Paddison_2013}. Analysis of the pNPD data was performed with the open source software cryspy \cite{cryspy}.

\section{Results and Discussion}

\subsection{Structural analysis of Ho$_2$TiO$_5$}

\begin{figure}[!ht]
    \centering
    \includegraphics[width=70mm]{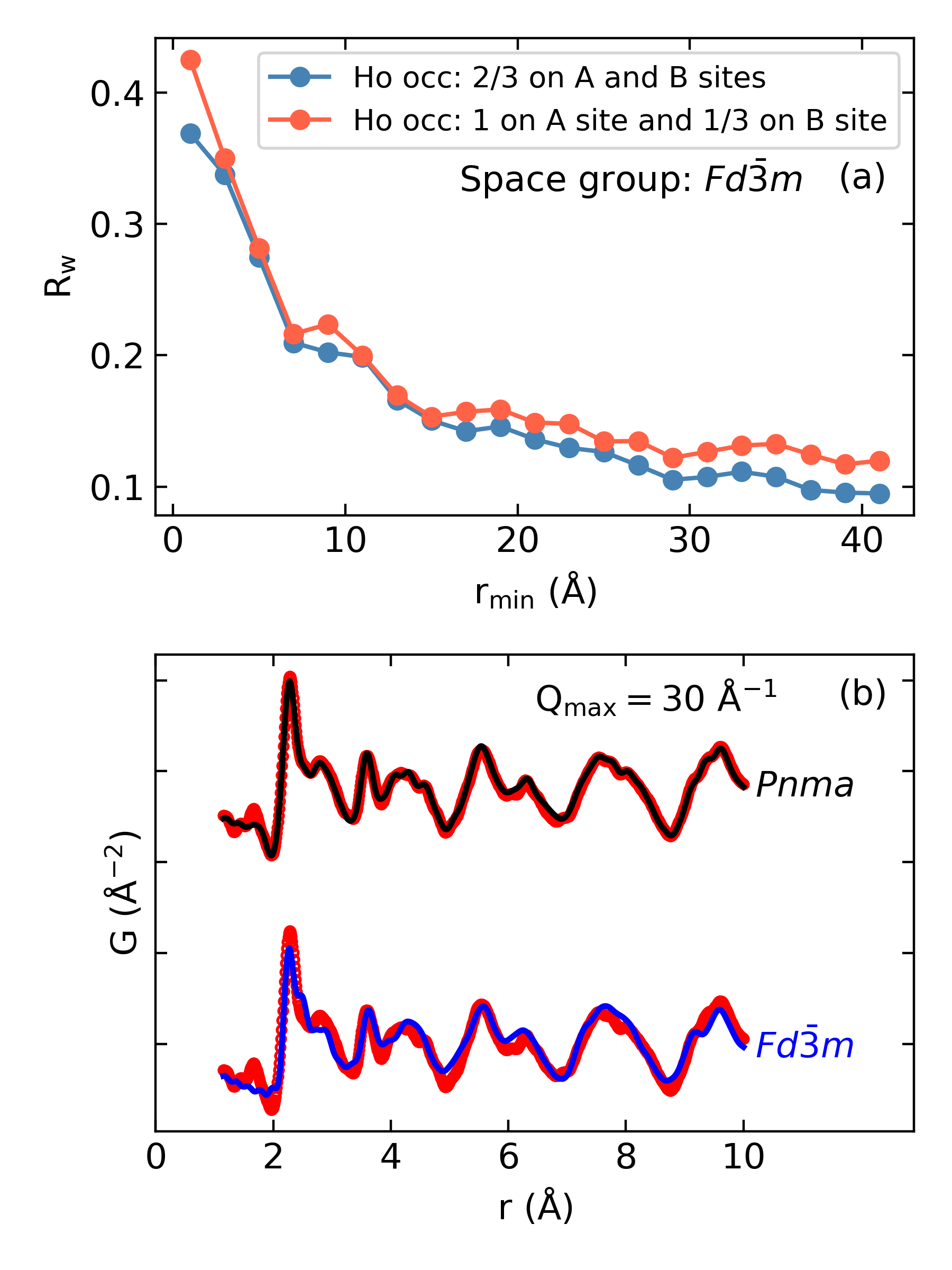}
    \caption{\label{fig:PDF_fit_Ho215} 
        (a) Goodness-of-fit parameter ($R_w$) as a function of minimum $r$ value ($r_\mathrm{min}$) in the “\textit{sliding-box}” refinements at 300 K. The blue symbol represent the Ho occupancy with 2/3 on both A and B sites, and the orange symbol represent the Ho occupancy where the A site is fully occupied and the B site has 1/3 occupancy. (b) Local atomic fits at 300 K using the $Fd\bar{3}m$ and $Pnma$ models over 1–10~\AA. (Neutron scattering data collected on the POWGEN instrument.)
    }
\end{figure}

We begin by investigating the average and local atomic structures of Ho$_2$TiO$_5$ by utilizing PDF analysis of neutron total scattering data collected on the POWGEN instrument at 300 K. This is well above any expected magnetic correlations, allowing a focus on the solely atomic scattering. The average structural information can be obtained by fitting the PDF data at the high $r$-region (e.g., 40--65 $\mathrm{\AA}$), whereas information about the local structure or any local distortion, if present, can be obtained by analyzing the low $r$-region (e.g. 1--10 $\mathrm{\AA}$) of the PDF data. 

To examine how the local structure evolves toward the average crystallographic structure, we performed a “\textit{sliding-box}" refinement of the PDF data using two different structural models within the $Fd\bar{3}m$ space group: one with Ho occupancy 2/3 on both A and B sites, and another with Ho occupancy 1 on the A site and 1/3 on the B site. The Ti ion fills the remaining occupancy in both models. In this approach, the fitting range is systematically shifted to higher $r$ values while keeping the total window width constant. Each refinement was carried out by maintaining the fitting range of 20 $\mathrm{\AA}$, starting from 1--21 $\mathrm{\AA}$ for the first window, then 3--23 $\mathrm{\AA}$, and so on, until the final range of 41--61 $\mathrm{\AA}$. The goodness-of-fit parameter $R_w = \sqrt{ \frac{ \sum_i ( y_{i,\mathrm{obs}} - y_{i,\mathrm{calc}} )^2 }{ \sum_i ( y_{i,\mathrm{obs}} )^2 } }$ was calculated for each refinement using the $Fd\bar{3}m$ structure. The plot of goodness-of-fit parameter ($R_w$) with minimum $r$ value ($\mathrm{r_{min}}$) fitting range is shown in  Fig.~\ref{fig:PDF_fit_Ho215} (a). The $R_w$ parameter initially decreases as the fitting range slides to higher values of $r$ for both models. Beyond $r_{\min} = 27$ \AA, $R_w$ saturates at ${\approx}\,0.095$ for the model with Ho occupancy $2/3$ on both A and B sites, and at ${\approx}\,0.125$ for the model with Ho occupancy 1 on the A site and $1/3$ on the B site. This indicates that the average $Fd\bar{3}m$ structure better captures the high-$r$ region of the PDF data when the $2/3$ occupancy model is used. In contrast, both models fail to capture the low-$r$ region (fitting range $1$--$10$ \AA), revealing that the local structure deviates from the average crystallographic framework (see Fig.~\ref{fig:PDF_fit_Ho215}\,(b); fit shown in blue).

\begin{figure}[tb]
    \centering
    \includegraphics[width=80mm]{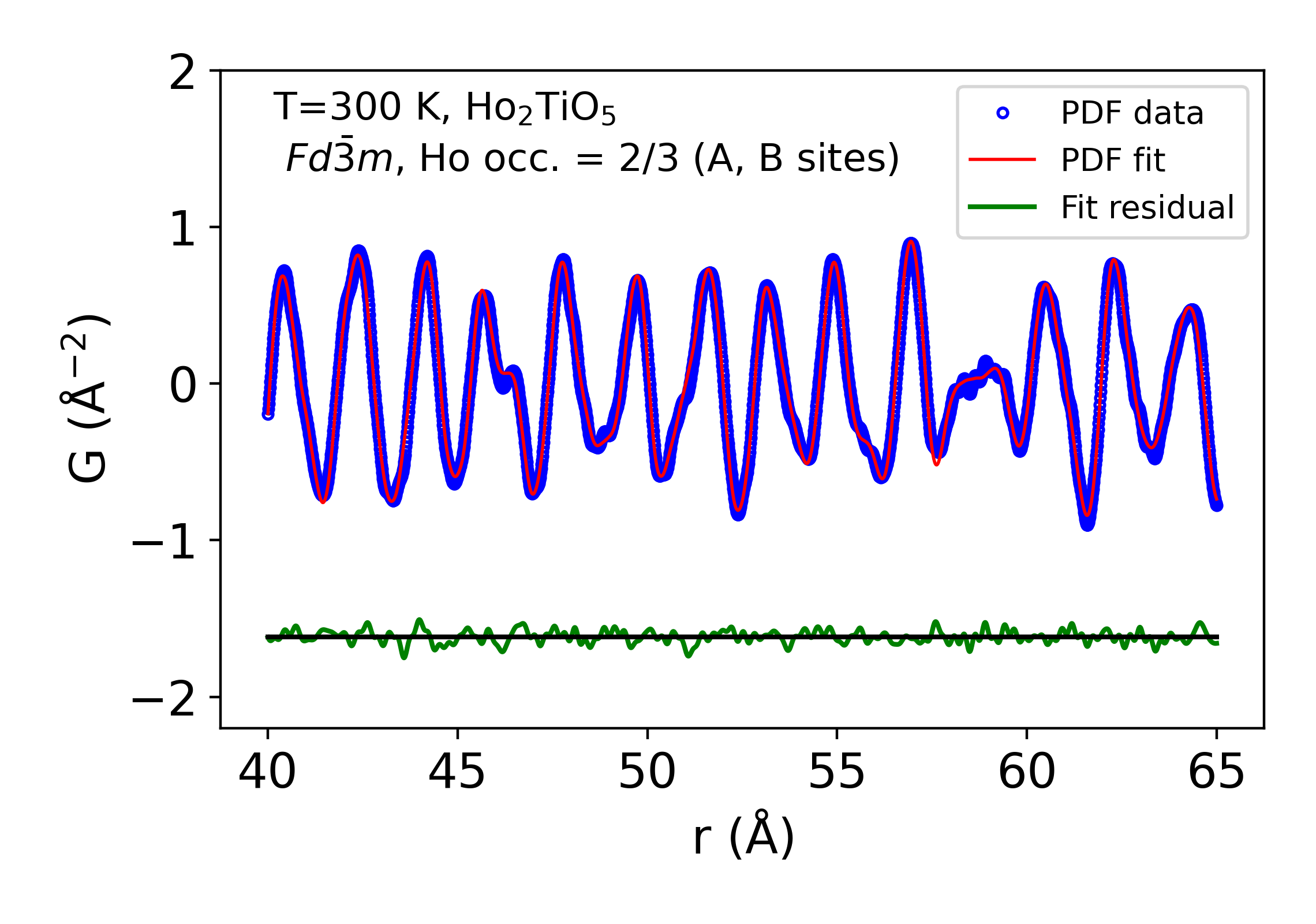}
    \caption{PDF pattern of Ho$_2$TiO$_5$ collected on POWGEN instrument at $T = 300$ K
    ($Q_\mathrm{max} = 30$ \AA$^{-1}$).}
    \label{fig:av_PDF_fit}
\end{figure}

The poor local structure fit motivated us to explore alternative structural models. The atomic PDF fit performed using the $Pnma$ space group well captured the local fit (1--10~\AA), suggesting that the true local structure deviates from the average $Fd\bar{3}m$ symmetry (see Fig.~\ref{fig:PDF_fit_Ho215}(b)). However, we caution that the adoption of $Pnma$ for the local atomic fit refinement introduces a substantially larger number of free parameters, raising the risk of overfitting. This concern is corroborated by the 10~K PDF data, where the $Pnma$ model produces an apparently good fit, with no residual magnetic signal in the fit. Inspection of the temperature-subtracted scattering, $S(Q,10~\mathrm{K}) - S(Q,50~\mathrm{K})$, reveals a clear diffuse magnetic signal that is suppressed when the low-symmetry $Pnma$ model is used. Instead, fitting to a lower-symmetry space group blends the atomic and magnetic contributions, thereby hiding the magnetic signal into the $Pnma$ model due to overfitting. Therefore, we do not proceed with using the $Pnma$ model for the local order and instead adopt the \(Fd\bar{3}m\) space group for our structural model and proceed with magnetic refinements based on this framework. The local misfits that persist at low $r$ are accordingly interpreted as a signature of short-range Ti--O, Ho--O and O--O bond disorder intrinsic to the stuffed pyrochlore lattice, rather than as evidence of a global symmetry breaking.

Fig.~\ref{fig:av_PDF_fit} shows the pair distribution function (PDF) fit of Ho$_2$TiO$_5$ at $T = 300$ K over the high-$r$ fitting range of 40--65 \AA, representing the average crystal structure. The data (blue symbols) are well reproduced by the PDF fit (red line) using the pyrochlore-type $Fd\bar{3}m$ structure with a fractional Ho occupancy of 2/3 on both the A and B sites. The flat and featureless fit residual (green line) confirms the excellent agreement between the observed and calculated PDF patterns. The quality of the fit at high $r$ demonstrates that the $Fd\bar{3}m$ average structure model 
successfully captures the long-range structural correlations in Ho$_2$TiO$_5$. Details of the crystal structure are given in Table~\ref{tab:cryst}.

\begin{table}[h]
\centering
\caption{Structural parameters for Ho$_2$TiO$_5$ extracted using POWGEN PDF data at 300 K using fitting range 40-65 $\mathrm{\AA}$ and the space group $Fd\bar{3}m$ (no.\ 227).
($U_{\rm iso}$\,=\,isothermal temperature factor; Occ\,=\,occupancy).}
\label{tab:cryst}
\begin{ruledtabular}
\begin{tabular}{llllllc}
\multicolumn{7}{l}{Lattice parameter: $a = 10.3008(24)$\,\AA} \\
\hline
Atom & \begin{tabular}[c]{@{}l@{}}Wyckoff\\position\end{tabular}
 & $x$ & $y$ & $z$ & $U_{\rm iso}$ & Occ \\
\hline
Ho(1) & $16c$ & 0          & 0       & 0       & 0.0454(14) & 0.667 \\
Ti(1) & $16c$ & 0          & 0       & 0       & 0.0128(19) & 0.333 \\
Ho(2) & $16d$ & 0.5        & 0.5     & 0.5     & 0.0146(4)  & 0.667 \\
Ti(2) & $16d$ & 0.5        & 0.5     & 0.5     & 0.0036(8)  & 0.333 \\
O(1)  & $8b$  & 0.375      & 0.375   & 0.375   & 0.0420(13) & 1.000 \\
O(2)  & $8a$  & 0.125      & 0.125   & 0.125   & 0.0760(47) & 0.730 \\
O(3)  & $48f$ & 0.3545(4) & 0.125   & 0.125   & 0.0665(11) & 0.820 \\
\end{tabular}
\end{ruledtabular}
\end{table}

\begin{figure}[!ht]
    \centering
    \includegraphics[width=80mm]{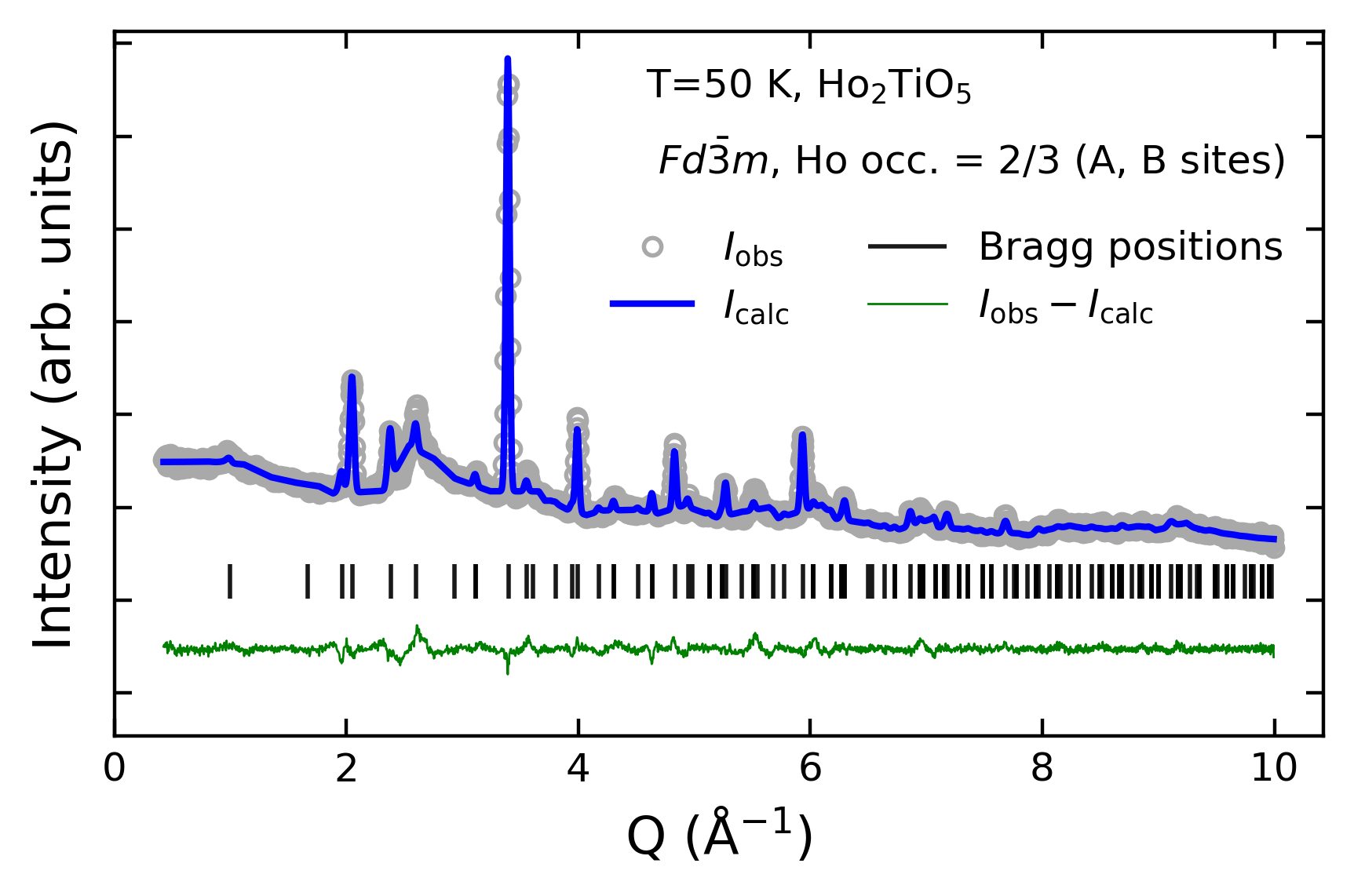}
    \caption{ 
    Rietveld refinement of the neutron powder diffraction pattern of Ho$_2$TiO$_5$ collected at 50 K on the HB2A instrument ($\lambda = 1.12\,\text{\AA}$).  
    }
    \label{fig:rietveld fit}
\end{figure}

The average crystal structure of Ho$_2$TiO$_5$, determined from total scattering data collected at POWGEN, is the pyrochlore-type \(Fd\bar{3}m\)  space group with Ho occupying both the A and B sites at 2/3 fractional occupancy, as shown in Fig.~\ref{fig:struc} (b). This result is further supported by Rietveld refinement of data collected at HB-2A, which confirms mixed A- and B-site occupancy and rules out fully occupied A sites with 1/3 occupancy on the B site.  A signature of the mixed occupancy can be seen by the (331) reflection being broad rather than sharp, as has been noted in previous studies \cite{LAU200845}.
Fig.~\ref{fig:rietveld fit} shows the Rietveld refinement of Ho$_2$TiO$_5$ performed using the \textsc{FullProf} suite~\cite{RODRIGUEZCARVAJAL199355_fullprofsuite} at $T = 50$ K. The refinement was carried out in the pyrochlore-type $Fd\bar{3}m$ space group with a fractional Ho occupancy of 2/3 on both the Ho1 (A site) and Ho2 (B site). The background was modeled by manually selecting points in the refinement, with particular attention given to the region between 2.41 and 2.9 \AA$^{-1}$ 
to accurately account for the atomic diffuse scattering contribution in this range. The Bragg peak positions are indicated by the black tick marks. The refinement converged with a goodness-of-fit of $\chi^2 = 2.66$, confirming the reliability of the structural model.

\subsection{Local site susceptibility measurements of Ho$_2$TiO$_5$}

\begin{figure}[htbp]
    \centering
    \includegraphics[width=\linewidth]{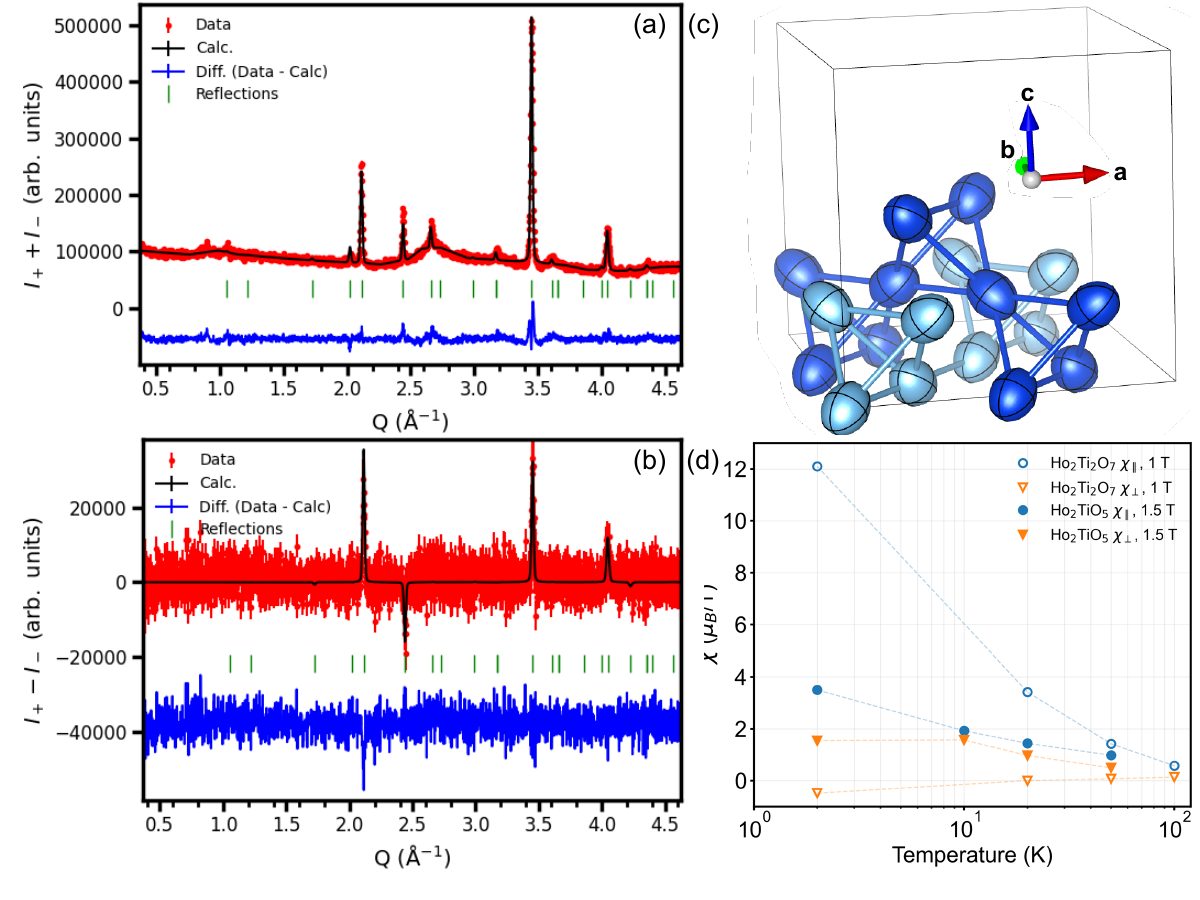}
    \caption{\label{fig:pNPD_Ho215} 
Half-polarized neutron powder diffraction for the (a) sum and (b) difference of $I_+$ and $I_-$ measurements of Ho$_2$TiO$_5$ at 10~K and $H=0.5$~T. (c) Local susceptibility ellipsoids for the Ho ions extracted from the model fits. (d) Parallel ($\chi_{\parallel}$) and perpendicular ($\chi_{\perp}$) components of the susceptibility as a function of temperature for Ho$_2$TiO$_5$ at $H=1.5$~T compared to Ho$_2$Ti$_2$O$_7$ at $H=1$~T. (Neutron scattering data collected on the HB-2A instrument).
    }
\end{figure}

To investigate the local magnetic anisotropy in Ho$_2$TiO$_5$, pNPD measurements were performed. This technique, described in detail in Ref.~\onlinecite{PhysRevResearch.1.033100}, uses two measurements with the incident neutron beam polarization parallel ($I_+$) or antiparallel ($I_-$) to an applied magnetic field at the sample position. Data were collected at 10~K in an applied field of 0.5~T, within the linear $M/H$ regime~\cite{PhysRevB.76.054430}. Both datasets were modeled using the \textsc{CrysPy} software~\cite{cryspy} to extract the local site susceptibility tensor, see Fig.~\ref{fig:pNPD_Ho215}(a)--(b). The Ho ions were assumed to be randomly distributed on both Ho1/Ti1 and Ho2/Ti2 sites with 2/3 occupancy on each.

A key question for the stuffed pyrochlore Ho$_2$TiO$_5$ is whether the B-site Ho$^{3+}$ ions, which sit in a 6-fold octahedral oxygen environment rather than the 8-fold coordination of the A site, retain local $\langle 111 \rangle$ Ising anisotropy. To address this, models were tested with both independent (unconstrained) and equal (Ho1=Ho2) susceptibility tensors. In the unconstrained refinement, both sites independently exhibit uniaxial anisotropy with the easy axis along the local $\langle 111 \rangle$ direction and quantitatively similar principal values: $\chi_{\parallel} = 3.80$~$\mu_\mathrm{B}$/T and $\chi_{\perp} = 1.82$~$\mu_\mathrm{B}$/T for the A-site Ho1, and $\chi_{\parallel} = 4.08$~$\mu_\mathrm{B}$/T and $\chi_{\perp} = 1.85$~$\mu_\mathrm{B}$/T for the B-site Ho2. Despite the distinct crystallographic coordination environments, the B-site Ho moments are not magnetically inert or differently oriented; they carry comparable $\langle 111 \rangle$ Ising anisotropy to the A-site Ho. This provides experimental evidence that the spin correlations develop along the local $\langle 111 \rangle$ directions across both magnetic sublattices.

Given the close agreement between the two sites, the constrained model (Ho1=Ho2) was adopted for final quantitative analysis. The resulting susceptibility tensor at 10~K and 0.5~T is:
\[
\chi_{\mathrm{Ho}} =
\begin{pmatrix}
2.29(16) & 0.65(38) & 0.65(38) \\
0.65(38) & 2.29(16) & 0.65(38) \\
0.65(38) & 0.65(38) & 2.29(16)
\end{pmatrix}
\, \mu_\mathrm{B}/\mathrm{T}
\]
The corresponding susceptibility ellipsoids are shown in Fig.~\ref{fig:pNPD_Ho215}(c). The easy axis points into or out of the center of the tetrahedra for both the Ho1 and Ho2 sublattices, confirming that the single-ion anisotropy required for spin-ice physics is preserved across the entire Ho network. Although the $\langle 111 \rangle$ easy axis is retained, the anisotropy ratio $\chi_{\parallel}/\chi_{\perp} \approx 2.1$ is reduced from the sharp Ising limit observed in Ho$_2$Ti$_2$O$_7$, consistent with a larger transverse susceptibility arising from the modified crystal-field environment due to cation mixing and oxygen vacancies in the stuffed lattice.

To further quantify how disorder affects the anisotropy, we compare the temperature-dependent $\chi_{\parallel}$ and $\chi_{\perp}$ values for Ho$_2$TiO$_5$ with those reported for Ho$_2$Ti$_2$O$_7$ in Ref.~\onlinecite{mPDF_HB2A_Baral} (Fig.~\ref{fig:pNPD_Ho215}(d)). At low temperature, clear anisotropy is present in both compounds, with the easy axis promoting spins to point into or out of the tetrahedra. However, this anisotropy is damped significantly more rapidly with increasing temperature in Ho$_2$TiO$_5$ than in Ho$_2$Ti$_2$O$_7$, analogous to how thermal fluctuations soften the Ising character in the parent compound, but occurring at a lower energy scale due to the structural disorder. The retention of $\langle 111 \rangle$ Ising anisotropy on both cation sublattices establishes that the single-ion prerequisite for spin-ice correlations survives the stuffing process, and that the disruption of ice-rule order in Ho$_2$TiO$_5$ arises from the incomplete tetrahedral connectivity rather than a loss of the underlying Ising character.

\begin{figure*}[htbp]
    \centering
     \includegraphics[width=140mm]{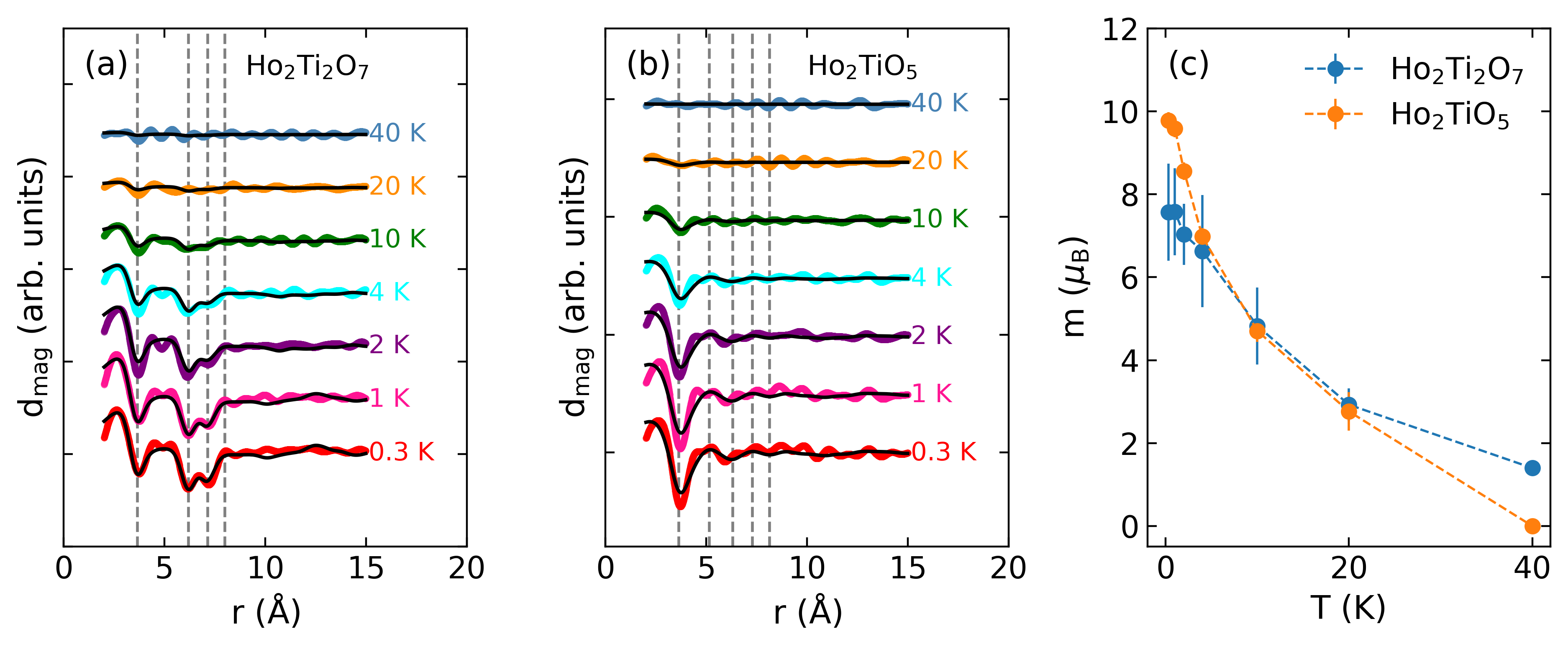}
    \caption{\label{fig:mPDF} 
 Temperature dependent mPDF fits of (a) Ho$_2$Ti$_2$O$_7$ and (b) Ho$_2$TiO$_5$. The mPDF patterns are vertically offset for clarity. A Fourier filter was applied to smooth the experimental signal, removing contribution above 6.5 $\mathrm{\AA^{-1}}$, this cutoff is justified by the magnetic form factor, which causes the magnetic scattering intensity to become neligble at higher $Q$. (c) Temperature dependence of the Ho mPDF moment amplitude extracted from mPDF fits for Ho$_2$Ti$_2$O$_7$ and Ho$_2$TiO$_5$. (Neutron scattering data collected on the HB-2A instrument). 
    }
\end{figure*}

\subsection{Magnetic pair distribution function analysis}

With an understanding of the anisotropy of the spins in both Ho$_2$Ti$_2$O$_7$ and Ho$_2$TiO$_5$ we now turn to probing the magnetic correlations, which remain short range down to the lowest temperatures measured in zero field in both materials. mPDF offers a route to investigate short range order in real space. We begin with Ho$_2$Ti$_2$O$_7$ as a reference, however, we note that direct analysis of mPDF models against data does not appear in the literature for Ho$_2$Ti$_2$O$_7$. We then extend our treatment of mPDF analysis to Ho$_2$TiO$_5$.

\subsubsection{mPDF analysis of Ho$_2$Ti$_2$O$_7$}

We analyzed the short-range magnetic correlations of Ho$_2$Ti$_2$O$_7$ in real space using the mPDF technique. The magnetic PDF signal was isolated by first performing an atomic PDF refinement using the established $Fd\bar{3}m$ crystal structure of Ho$_2$Ti$_2$O$_7$. The resulting fit residual, defined as the difference between the measured PDF and the calculated atomic PDF, was then subjected to temperature subtraction in real space using the atomic fit residual of 50 K data as a reference \cite{Braedon_2024}. The 50 K temperature was chosen as the reference because any magnetic correlations at this temperature are negligible, ensuring that the subtraction removes only nuclear scattering contributions and artifacts without affecting the magnetic signal of interest. The resulting clean mPDF signal was subsequently used as input for the mPDF refinement.

Fig.~\ref{fig:mPDF}(a) shows the mPDF refinement of Ho$_2$Ti$_2$O$_7$ at various temperatures. At 0.3 K, a pronounced negative peak is observed at the nearest-neighbor distance of $\approx 3.57\,\text{\AA}$, reflecting the expected nearest-neighbor correlation for spin-in/spin-out configurations on a tetrahedron in the global spin basis. Two additional negative peaks appear at $\approx 6.18\,\text{\AA}$  and $\approx 7.14\,\text{\AA}$, corresponding to the second and third nearest-neighbor distances, respectively. The magnetic model captures all three peaks well, though the agreement diminishes at larger distances as the peaks become progressively weaker. This behavior indicates that non-random spin correlations persist up to a length scale of approximately 7–8\,\AA\ at 0.3\,K.

In order to perform the mPDF refinement of Ho$_2$Ti$_2$O$_7$ a custom python script was used to generate a 2-in/2-out spin configuration. Thus created spin configuration was used to perform mPDF refinement and the local magnetic moment was calculated. As 2-in/2-out spin configuration can be realized through multiple distinct spin arrangement, the refinement was repeated over several independent configurations and the resulting local moment were averaged to obtain a statistically robust value.

\subsubsection{mPDF analysis of Ho$_2$TiO$_5$}

Having established the crystal structure and local magnetic anisotropy of Ho$_2$TiO$_5$ above, and with the example of Ho$_2$Ti$_2$O$_7$, we now turn to the analysis of the short range magnetic correlations at low temperature directly in real space using the mPDF technique for Ho$_2$TiO$_5$. The isolation of the magnetic PDF signal was done  in a similar manner to Ho$_2$Ti$_2$O$_7$ using $Fd\bar{3}m$ space group with Ho occupancy 2/3 on both A and B sites. The atomic PDF fits were performed at the temperature ranging from 0.3-50 K. The atomic fit residual of the 50 K was subtracted to all other temperature in a real space \cite{Braedon_2024} to get a clean mPDF signal which was used as input for the mPDF refinement. Ho$_2$Ti$_2$O$_7$ was first used as a reference to establish the mPDF refinement, where Ho site (Ho on A site) follows a perfect 2-in-2-out spin-ice configuration. Since Ho$_2$TiO$_5$ has a fractional occupancy of 2/3 on both the A and B sites, the 2-in-2-out model was extended to include both sites, using the Ho$_2$Ti$_2$O$_7$ refinement as the starting point. 

Fig.~\ref{fig:mPDF}(b) shows the mPDF refinements of Ho$_2$TiO$_5$ at various temperatures using the spin-ice structural model. At 0.3 K, a pronounced  negative nearest-neighbor mPDF peak at $\approx 3.6\,\text{\AA}$ reflects the expected relative orientation of neighboring Ho moments in the local spin-in/spin-out configurations on a tetrahedron. Because the local $\langle111\rangle$ axes are noncollinear, this real-space correlation should not be interpreted simply as conventional collinear antiferromagnetism, but rather as the mPDF signature of local spin-ice-like correlations. The second-neighbor correlation produces a positive peak at $\approx 5.14\,\text{\AA}$, while the third-neighbor correlation produces a negative peak at $\approx 6.3\,\text{\AA}$. The magnetic model captures all three peaks well, although the agreement diminishes at larger distances as the peaks become progressively weaker. This behavior indicates that non-random spin correlations persist over a length scale of approximately 6--7~\AA\ at 0.3~K.

A series of mPDF fits were performed for both compositions, Ho$_2$Ti$_2$O$_7$ and Ho$_2$TiO$_5$  (see Fig.\ref{fig:mPDF} (a, b), across a temperature range of 0.3--40~K. The fits were performed over 2--15~\AA, and the refined magnetic mPDF amplitude was used as an effective correlated moment that tracks the strength of the local magnetic correlations. The temperature dependence of this effective moment is shown in Fig.~\ref{fig:mPDF}(c). At 0.3~K, the refined effective moment is approximately 7.57~$\mu_\mathrm{B}$ for Ho$_2$Ti$_2$O$_7$ and 9.7~$\mu_\mathrm{B}$ for Ho$_2$TiO$_5$. We do not interpret this difference as a change in the ionic Ho$^{3+}$ moment. Rather, it reflects differences in the effective mPDF scale arising from the different local structural models, Ho occupancies, and finite-range correlated spin configurations used for the two compounds. 

With increasing temperature, the effective correlated moment decreases monotonically in both compositions, consistent with thermal disordering of the local spin configurations. Notably, at 40~K the effective moment in Ho$_2$Ti$_2$O$_7$ remains finite at ${\sim}1.47~\mu_\mathrm{B}$, whereas that of Ho$_2$TiO$_5$ approaches zero, indicating the absence of detectable short-range magnetic correlations above this temperature within the sensitivity of the present mPDF analysis. This difference in energy scale is consistent with the stronger disruption of spin-ice correlations by Ho/vacancy disorder in Ho$_2$TiO$_5$.

\section{Reverse Monte Carlo modeling of diffuse magnetic scattering}

To form a complete picture of the microscopic magnetism in Ho$_2$TiO$_5$ we extend our investigation by combining both reciprocal-space and real-space analysis methods, as has been shown to be advantageous in studies of short-range magnetic order in frustrated magnets~\cite{Nutall2023}. In reciprocal space the Reverse Monte Carlo (RMC) technique was implemented in \textsc{Spinvert} to analyze the broad diffuse magnetic scattering that is observed with decreasing temperature in Ho$_2$TiO$_5$ and Ho$_2$Ti$_2$O$_7$, without any transition to long range magnetic order. This methodology allows for a model independent testing of spin arrangements, both within the anisotropic constraints of spins pointing in/out of tetrahedra and for unconstrained spins. We use Ho$_2$Ti$_2$O$_7$ as a reference and extend this to Ho$_2$TiO$_5$.

\subsubsection{RMC modeling of Ho$_2$Ti$_2$O$_7$}

\begin{figure}[!ht]
    \centering
    \includegraphics[width=85mm]{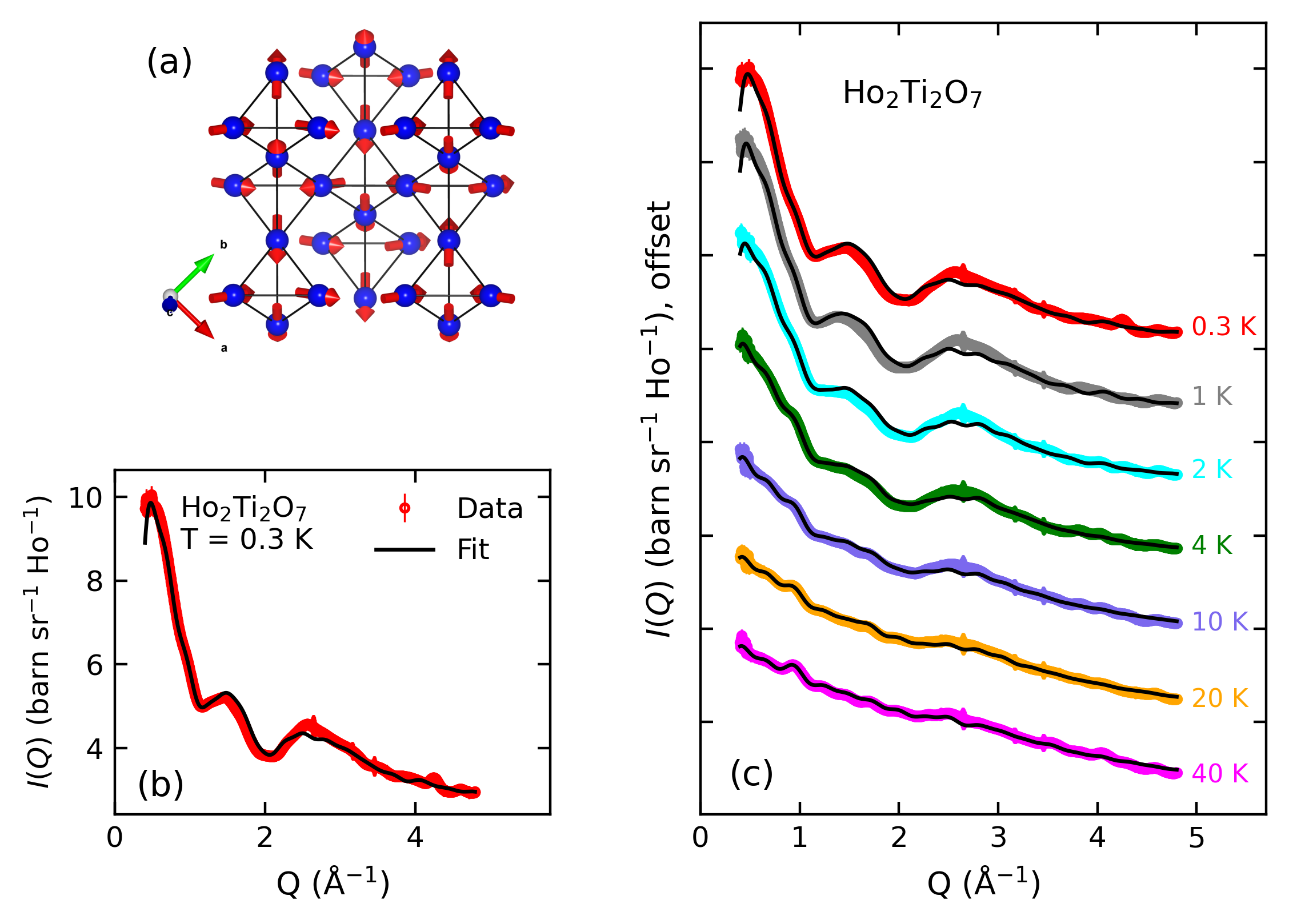}
    \caption{\label{fig:Ho227_Tseries} 
    (a) 2-in 2-out spin structure. (b) Reverse Monte Carlo (RMC) modeling of the magnetic diffuse scattering of Ho$_2$Ti$_2$O$_7$ at 0.3 K using \textsc{Spinvert}. The red symbols represent the experimental magnetic diffuse scattering data, while the black line shows the \textsc{Spinvert} fit. (c) Vertically stacked temperature series of magnetic diffuse scattering data (colored curves) in $Q$-space with their respective \textsc{Spinvert} fit (black line). (Neutron scattering data collected on the HB-2A instrument).
    }
\end{figure}

\begin{figure}
    \centering
    \includegraphics[width=\linewidth]{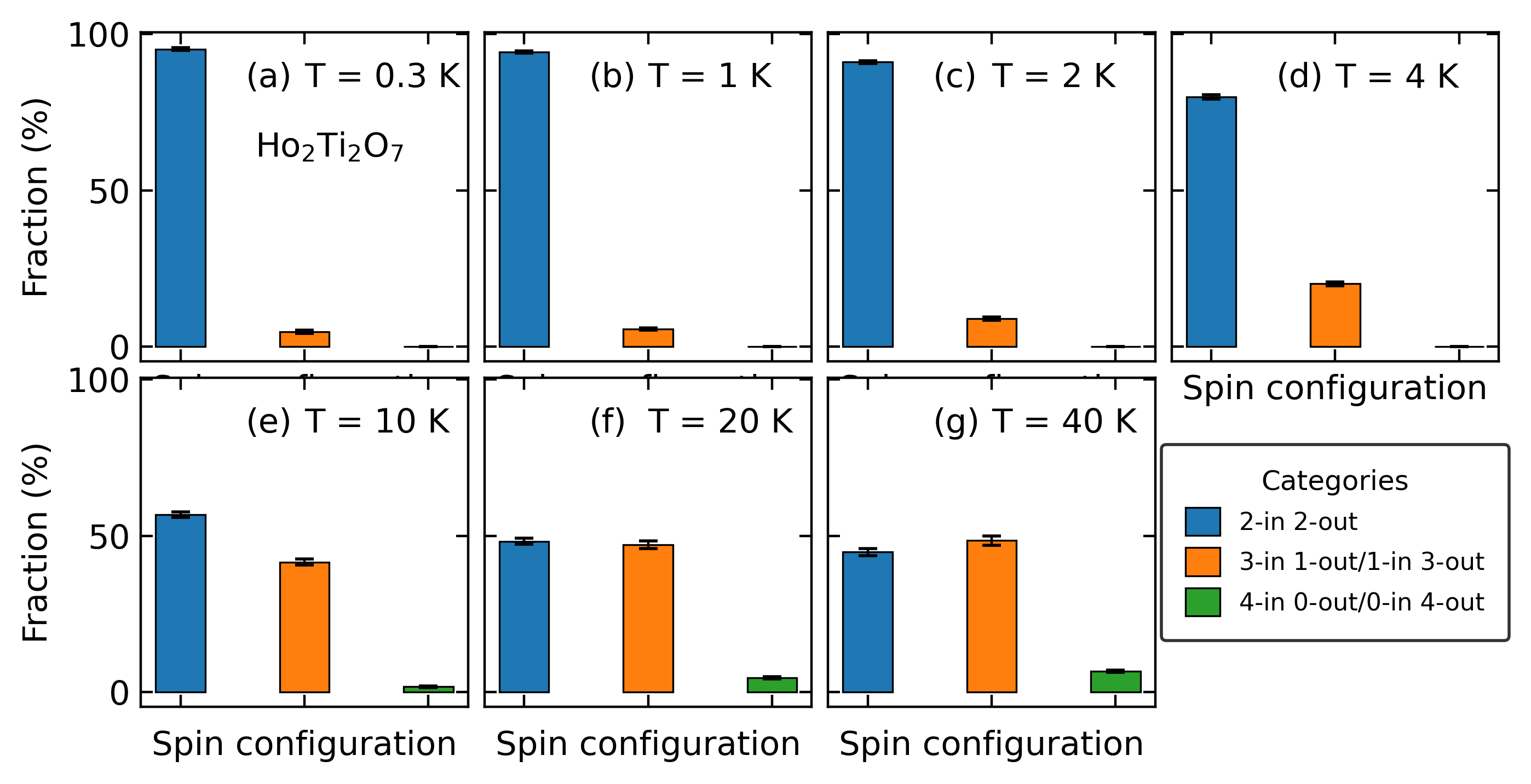}
    \caption{\label{fig:Ho227_fraction_bar} 
Bar charts showing the fraction of tetrahedra adopting each spin configuration as a function of temperature for Ho$_2$Ti$_2$O$_7$ obtained from RMC refinements. Blue, orange, and green bars represent the 2-in 2-out, 3-in 1-out/1-in 3-out, and 4-in 0-out/0-in 4-out spin configurations, respectively. Error bars denote the standard deviation over 100 independent RMC refinements. (Neutron scattering data collected on the HB-2A instrument).
    }
\end{figure}

Prior to RMC modeling of the magnetic scattering in the neutron diffraction data of Ho$_2$Ti$_2$O$_7$, the magnetic signal was isolated from the nuclear contribution. Bragg peaks were identified and removed from the powder dataset using an automated Python algorithm~\cite{mPDF_HB2A_ZnMnTe}, leaving only the diffuse magnetic scattering component. Thus extracted diffuse magnetic scattering components were subsequently normalized to absolute intensity~\cite{Paddison_normalization} and then RMC refinements were carried out using \textsc{Spinvert} software. For RMC calculations, we defined a $5 \times 5 \times 5$ supercell, and 100 independent RMC refinements were performed and averaged to improve statistical reliability.

The RMC fit for Ho$_2$Ti$_2$O$_7$ at 0.3~K is shown in Fig.~\ref{fig:Ho227_Tseries} (b). The RMC fit  demonstrates good agreement between the model and the data across the measured $Q$-range. Analysis of the refined spin structure of Ho$_2$Ti$_2$O$_7$ at 0.3 K reveals that approximately 95\% of tetrahedrah satisfy the 2-in 2-out ice rule, while the remaining 5\% of tetrahedral are following 3 in 1 out or 1 in 3 out spin arrangement. Similar RMC refinements were performed across the full temperature range from 0.3 to 50~K. Fig~\ref{fig:Ho227_Tseries}(c) shows the \textsc{Spinvert} fit of whole temperature series, and resulting spin configuration fractions are summarized in Fig.~\ref{fig:Ho227_fraction_bar}. The ice rule compliance remains dominant at low temperatures, with approximately 95\% of tetrahedra satisfying the 2-in 2-out condition at 0.3, 1, and 2~K. As the temperature increases, the fraction of ice-rule-obeying tetrahedra decreases monotonically, accompanied by a corresponding increase in the 3-in 1-out/1-in 3-out and all-in all-out population. At 20~K and above, the two spin configurations (2-in 2-out and 3-in 1-out/1-in 3-out) are present in nearly equal proportions, indicating a progressive breakdown of spin ice correlations with increasing thermal fluctuations (see Fig.~\ref{fig:Ho227_fraction_bar})

\subsubsection{RMC modeling of Ho$_2$TiO$_5$}

\begin{figure*}[!ht]
    \centering
    \includegraphics[width=170mm]{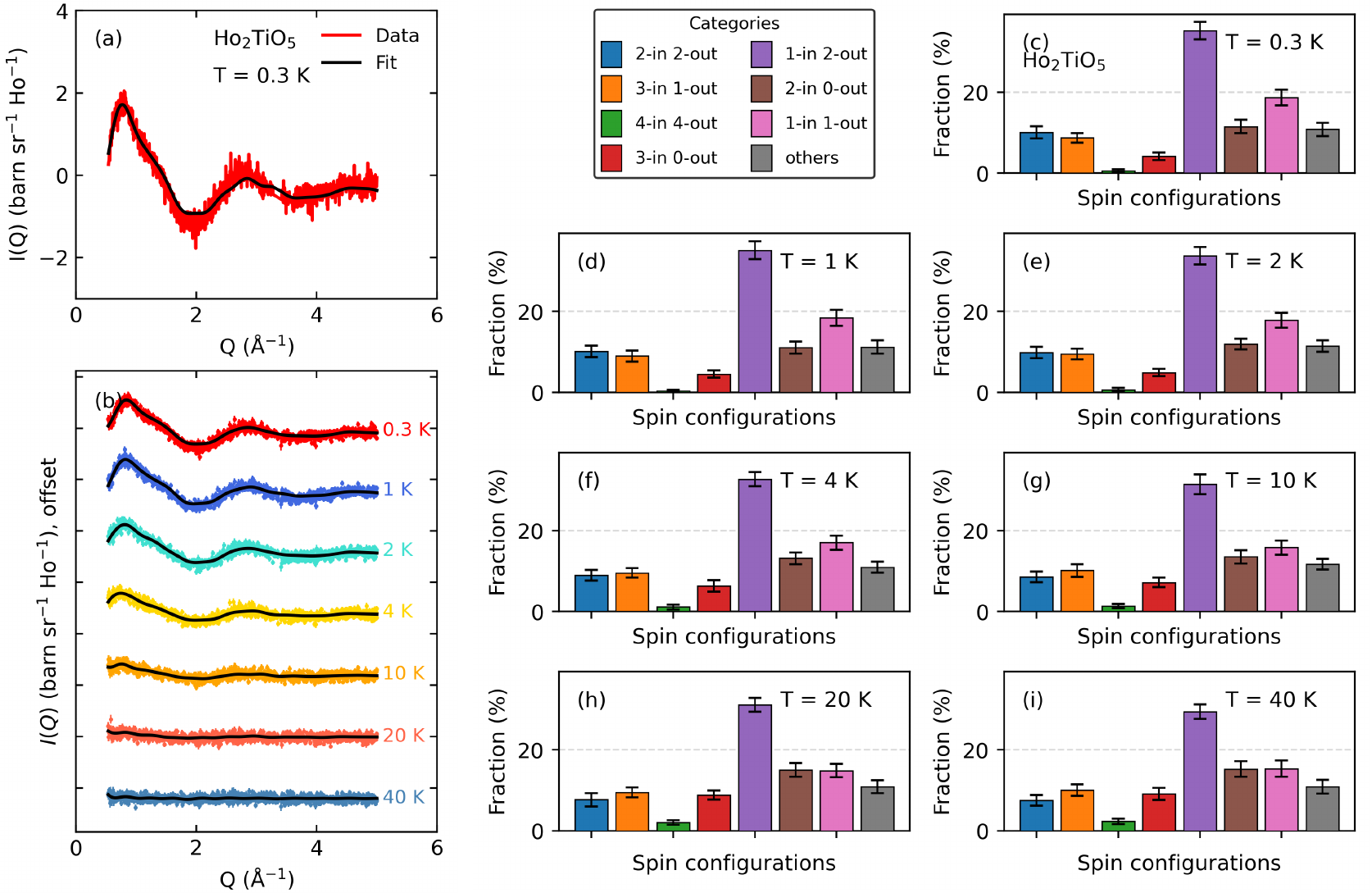}
    \caption{\label{fig:spinvertfit_Ho215} 
(a) RMC fit to the magnetic diffuse scattering of Ho$_2$TiO$_5$ at 0.3 K. Red symbols show the data and the black line shows the fit. (b) Temperature series of RMC fits, vertically offset for clarity. (c)--(i) Fractions of tetrahedral spin configurations extracted from the RMC refinements as a function of temperature. Colors denote 2-in/2-out, 3-in/1-out or 1-in/3-out, 4-in/0-out or 0-in/4-out, 3-in/0-out or 0-in/3-out, 2-in/1-out or 1-in/2-out, 2-in/0-out or 0-in/2-out, 1-in/1-out, and single-spin configurations. (Neutron scattering data collected on the HB-2A instrument).}

\end{figure*}

To investigate the nature of the magnetic diffuse scattering and its evolution with temperature, RMC refinements were performed against the total neutron scattering data of polycrystalline Ho$_2$TiO$_5$ collected at temperatures ranging from 0.3 K - 50 K. To isolate the diffuse magnetic signal a temperature subtraction approach was employed. The 50 K dataset, where magnetic correlations are negligible, was used as a reference and subtracted from all lower-temperature datasets. The resulting difference patterns contain the isolated magnetic diffuse scattering, which was then modeled using \textsc{Spinvert}.

The magnetic RMC analysis of Ho$_2$TiO$_5$ was carried out using \textsc{Spinvert} based on the average $Fd\bar{3}m$ structure, employing a $5 \times 5 \times 5$ supercell with fractional occupancies of 2/3 for both the Ho1 (A site) and Ho2 (B site). Vacancies were introduced by randomly removing Ho atoms from both sites to match the 2/3 occupancy, and this process was repeated 100 times with different random vacancy configurations to ensure a well-converged statistical ensemble. Two different refinement approaches were performed; spins constrained along the local $\langle 111 \rangle$ directions (into or out of the Ho$^{3+}$ tetrahedra), and spins freely rotating in three dimensions. While both yielded comparable fit quality, the unconstrained model produced randomly oriented spins with no physically meaningful correlations, indicative of overfitting. All subsequent analyses were therefore performed with the $\langle 111 \rangle$ Ising constraint, consistent with the established single-ion anisotropy of Ho$^{3+}$ in the pyrochlore environment and the pNPD results presented above for the local anisotropy.

Fig.~\ref{fig:spinvertfit_Ho215}(a) shows the \textsc{Spinvert} fit of the 0.3 K data of Ho$_2$TiO$_5$, demonstrating excellent agreement between the RMC model and the experimental diffuse scattering across the measured $Q$-range. Fig.~\ref{fig:spinvertfit_Ho215}(b) presents the \textsc{Spinvert} fits at all measured temperatures, where the data are vertically offset for clarity. The quality of the fits across the full temperature range confirms the reliability of the extracted spin configurations and allows a systematic investigation of how the diffuse magnetic correlations evolve with temperature in this material.

The RMC modeling of Ho$_2$TiO$_5$ at 0.3 K data reveals distinctions when contrasted against the canonical spin-ice behavior observed in Ho$_2$Ti$_2$O$_7$. Fig.~\ref{fig:spinvertfit_Ho215}(c-i) show the fraction of spin configuration for the temperature series. Approximately 20\% of tetrahedra are magnetically complete, in agreement with the value $(2/3)^4 \approx 0.20$ expected for random $2/3$ occupancy of the four tetrahedral vertices, while the remaining ${\sim}80\%$ contain at least one vacancy. Out of 20\% complete tetrahedra only 10\%  follow the classical 2-in 2-out spin configuration whereas 9\% follow the 3-in 1-out or 1-in 3-out spin configuration, and about 1\% follow an all-in/all-out spin configuration. This stands in a sharp contrast to Ho$_2$Ti$_2$O$_7$ where the 2-in 2-out spin configuration accounts for $\sim$95\% of tetrahedra, indicating that the spin-ice correlations are significantly disrupted in Ho$_2$TiO$_5$.

A large population (80\%) of tetrahedra are partially occupied in Ho$_2$TiO$_5$. Among those the majority of the tetrahedra have 1-in 2-out or 2-in 1-out spin configurations. These configurations are ice-rule-compatible in the sense that restoring the missing Ho spin with the appropriate orientation would produce a 2-in/2-out tetrahedron. This suggests that a significant fraction of the Ho$_2$TiO$_5$ spin configurations retain a local tendency toward ice-rule compliance, despite the structural disorder introduced by site vacancies. The rest are 3-in 0-out or vice-versa, 2-in 0-out or vice-versa, 1-in 1-out, and single spin tetrahedra (See Fig.~\ref{fig:spinvertfit_Ho215}(c)). These results indicate that Ho$_2$TiO$_5$ does not exhibit fully well-defined spin-ice order characteristics compared to Ho$_2$Ti$_2$O$_7$, but instead the local spin correlations retain a residual ice-rule tendency that is frustrated by the structural vacancies on the magnetic sublattice. The system can therefore be described as a \textit{vacancy-disrupted spin-ice}, in which the geometric frustration inherent to the underlying tetrahedral motif competes with the disorder introduced by the non-stoichiometric holmium site occupancy. 

Unlike Ho$_2$Ti$_2$O$_7$, where warming produces a pronounced transfer of population from 2-in/2-out configurations into 3-in/1-out and 1-in/3-out configurations, the configuration fractions in Ho$_2$TiO$_5$ show only weak temperature dependence over the measured range. This suggests that the dominant limitation on ice-rule order in Ho$_2$TiO$_5$ is not thermal population of monopole-like defects, but the quenched Ho/Ti disorder that fixes a large fraction of the tetrahedra in magnetically incomplete configurations. The temperature dependence is instead more clearly reflected in the weakening of the diffuse-scattering intensity and the mPDF amplitude.

\section*{Conclusion}

In conclusion, we have combined neutron total scattering, mPDF analysis, reciprocal-space RMC modeling, and half-polarized neutron powder diffraction to determine how local spin-ice correlations evolve from the classical pyrochlore spin ice Ho$_2$Ti$_2$O$_7$ to the vacancy-disrupted stuffed compound Ho$_2$TiO$_5$. Both compounds retain $Fd\bar{3}m$ symmetry at the average structural level. In Ho$_2$TiO$_5$, however, this average description does not fully capture the local structure. The low-$r$ PDF misfit is consistent with short-range Ho--O, Ti--O and O--O bond-length disorder associated with partial Ho/vacancy occupancy, rather than a true symmetry-lowering transition.

Despite this structural disorder, the magnetic correlations in Ho$_2$TiO$_5$ retain a clear local spin-ice tendency. The diffuse scattering and mPDF results show that the Ho moments retain local spin-in/spin-out correlations on the tetrahedral network. This interpretation is supported by half-polarized neutron powder diffraction, which shows that the local $\langle 111\rangle$ Ising anisotropy is preserved even in the disordered stuffed lattice. The anisotropy is broadened relative to the sharp Ising limit expected for ideal Ho$_2$Ti$_2$O$_7$, consistent with a distribution of local crystal-field environments around the Ho ions.

For the reference spin ice Ho$_2$Ti$_2$O$_7$, RMC and mPDF analyses reveal a nearly ideal two-in/two-out state at base temperature, with approximately 95\% of tetrahedra obeying the spin-ice rule. Upon warming, this local ice-rule order progressively weakens, as an increasing fraction of tetrahedra adopt monopole-like three-in/one-out or one-in/three-out configurations. This behavior provides a benchmark for identifying the effects of vacancy disorder in Ho$_2$TiO$_5$.

In Ho$_2$TiO$_5$, the partial Ho occupancy produces a topologically incomplete magnetic network in which most tetrahedra contain one or more vacant magnetic sites. Nevertheless, the dominant incomplete configurations are 2-in/1-out and 1-in/2-out, which are locally compatible with the spin-ice rule if the vacant site is restored. Thus, Ho$_2$TiO$_5$ is not simply a disorder-destroyed spin ice. Instead, it realizes a vacancy-disrupted spin ice in which the local ice-rule tendency survives on an incomplete tetrahedral network. These results show that the single-ion prerequisites for spin-ice physics can survive strong magnetic dilution while the collective ice-rule ordering is substantially degraded, providing a microscopic picture of how structural disorder reshapes frustration rather than simply destroying it.

\section*{Data availability}
The data that support the findings of this study are available from the corresponding authors upon reasonable request.

\section*{Acknowledgments}
This research used resources at the High Flux Isotope Reactor and Spallation Neutron Source, a DOE Office of Science User Facility operated by Oak Ridge National Laboratory. Beam time was allocated on HB-2A (POWDER) under proposals IPTS-31471 and IPTS-31582, and on POWGEN under proposal IPTS-34824. The sample synthesis (H. Z.) was supported by the U.S. Department of Energy (DOE) under Grant No. DE-SC0020254. B.A.F. was supported by the U.S. National Science Foundation, Division of Materials Research, LEAPS-MPS program through Award No. 2418438. We thank Joe Paddison for helpful discussions on magnetic RMC modeling.

\bibliography{ref}

\end{document}